\documentstyle[aps,preprint,tighten]{revtex}

\newcommand{\lsim}{\mathrel{\mathop{\kern 0pt \rlap
  {\raise.2ex\hbox{$<$}}}
  \lower.9ex\hbox{\kern-.190em $\sim$}}}
\newcommand{\gsim}{\mathrel{\mathop{\kern 0pt \rlap
  {\raise.2ex\hbox{$>$}}}
  \lower.9ex\hbox{\kern-.190em $\sim$}}}
\newcommand{\beq}    {\begin{equation}}
\newcommand{\eeq}    {\end{equation}}
\newcommand{\beqarr} {\begin{eqnarray}}
\newcommand{\eeqarr} {\end{eqnarray}}
\newcommand{\barr}   {\begin{array}}
\newcommand{\earr}   {\end{array}}

\begin{document}

\preprint{
\begin{tabular}{r}
DFTT 49/98
\end{tabular}
}

\title{Combining the data of annual modulation effect in WIMP 
direct detection with  measurements of WIMP indirect searches}

\author{
\bf A. Bottino$^{\mbox{a}}$
\footnote{E--mail: bottino@to.infn.it,donato@to.infn.it,
fornengo@to.infn.it,scopel@posta.unizar.es},
F. Donato$^{\mbox{a}}$, N. Fornengo$^{\mbox{a}}$, 
S. Scopel$^{\mbox{b}}$\footnote{INFN Post--doctoral Fellow}
\vspace{6mm}
}

\address{
\begin{tabular}{c}
$^{\mbox{a}}$
Dipartimento di Fisica Teorica, Universit\`a di Torino and \\
INFN, Sezione di Torino, Via P. Giuria 1, 10125 Torino, Italy
\\
$^{\mbox{b}}$ Instituto de F\'\i sica Nuclear y Altas Energ\'\i as, \\
Facultad de Ciencias, Universidad de Zaragoza, \\
Plaza de San Francisco s/n, 50009 Zaragoza, Spain
\end{tabular}
}
\maketitle

\begin{abstract}

In previous papers we showed that the data of the DAMA/NaI 
experiment for WIMP direct detection, which indicate a possible 
annual modulation effect, are widely compatible with an
interpretation in terms of a relic neutralino as the major component of dark
matter in the Universe. In the present note we discuss the 
detectability of the relevant supersymmetric neutralino 
configurations by two of the most promising methods of indirect 
search for relic particles: measurement of cosmic--ray antiprotons and 
measurement of neutrino fluxes from Earth and Sun.

\end{abstract}  

\vspace{1cm}


\section{Introduction}

In Refs. \cite{noi0,noi1,noi2} 
 we showed that the indication of  a possible 
annual modulation  effect in WIMP direct search \cite{dama1,dama2}
are interpretable 
in terms of a relic neutralino which may make up the major 
part of dark matter in the Universe. 

We recall that 
the DAMA/NaI data \cite{dama2}  single out 
a very delimited 2--$\sigma$ C.L. region in the plane 
$\xi \sigma_{\rm scalar}^{\rm (nucleon)}$ -- $m_\chi$, 
where $m_\chi$ is the WIMP mass, 
$\sigma_{\rm scalar}^{\rm (nucleon)}$ is the WIMP--nucleon  scalar elastic
cross section  and $\xi = \rho_\chi / \rho_l$ 
is the fractional amount of local 
WIMP density $\rho_\chi$ with respect to the total local 
dark matter density $\rho_l$. In Fig. 1 we display the region $R$, which is
obtained from the original 2--$\sigma$ 
C.L. region of Ref. \cite{dama2} (where $\rho_l$ is
normalized to the value $\rho_l = 0.3$ GeV cm$^{-3}$), by accounting for the 
uncertainty in the value of $\rho_l$. In obtaining region $R$ we have
conservatively considered the range: 
0.1 GeV cm$^{-3} \leq \rho_l \leq $ 0.7 GeV cm$^{-3}$, because of a possible 
flattening of the dark matter
halo  \cite{turner1,turner2}  and a possibly sizeable baryonic contribution to the
galactic dark matter \cite{turner}.
In Fig. 1 is also displayed 
the scatter plot of the supersymmetric configurations which, 
in our scanning of the susy parameter space,  turn out 
to be  contained in region $R$  (the set comprised of these 
configurations is called set $S$). 
The theoretical framework adopted here is the Minimal Supersymmetric 
extension of the Standard Model (MSSM), as illustrated in 
Refs. \cite{noi0,noi1}, to which we refer for details. 

In \cite{noi0,noi1,noi2} we have proved that a significant fraction of 
configurations of  
 set $S$ is explorable at accelerators in the near future.  
In the present paper, we discuss the detectability 
of these configurations by different means, i.e. by two of the most 
promising ways of indirect search for relic particles: 
measurement of 
cosmic--ray antiprotons \cite{pbar,pbarnostro,pierre} and 
measurement 
of neutrino fluxes from  Earth and  Sun \cite{flux,mosc,bere2}.

Preliminary discussions on the links between 
the possible annual modulation effect and the results of indirect 
searches have been presented in Refs. \cite{noi0,pierre}, in connection 
with  the former DAMA/NaI data (data collected in the running 
period \# 1)\cite{dama1}. Here we extend 
our investigation to the new results of the DAMA/NaI experiment 
(running period \#2)\cite{dama2}, 
which establish a further indication for the annual 
modulation effect and, combined with the former 
ones, single out an effect at 99.6\% C.L. 
In the present paper we also discuss neutralino 
cosmological properties in terms of its relic abundance 
$\Omega_{\chi} h^2$. This quantity is evaluated here as illustrated in 
\cite{ouromega}, and for the parameter $\xi$ we 
take $\xi = {\rm min}(\Omega_{\chi} h^2/(\Omega h^2)_{\rm min},1)$, 
with $(\Omega h^2)_{\rm min} = 0.01$. In exploring our supersymmetric 
parameter space we  apply an upper bound, conservatively set at the 
value $\Omega_\chi h^2 \leq 0.7$, and we consider as cosmologically 
interesting the range $0.01 \leq \Omega_{\chi} h^2 \leq 0.7$. However, 
we stress that, according to the most recent data and analyses 
\cite{omegamatter}, the most appealing interval for the 
neutralino relic abundance is 
$0.02 \lsim \Omega_{\chi} h^2 \lsim 0.2$. 

Most of our results for the various quantities considered in this paper 
will be provided in the form of scatter plots, obtained by varying 
the representative points of the susy parameter space over the 
configurations of set $S$. When in a specific figure (or in a panel of it) the 
value of 
$\rho_l$ is fixed, it is meant that the corresponding scatter plot 
concerns only the subset of  $S$ comprised of those 
configurations which belong to that particular value of $\rho_l$.  
For instance, for $\rho_l = 0.3$ GeV cm$^{-3}$, only the points inside 
the dashed curve of Fig. 1 are considered.

\section{Cosmic--ray antiprotons}

Relic neutralinos, if present in our halo, would produce antiprotons 
by  pair annihilation \cite{pbar,pbarnostro,pierre}. To
discriminate this potential source of primary cosmic--ray $\bar p$'s from the 
secondary antiprotons, i.e. $\bar{p}$'s
 created by interactions of primary cosmic--ray nuclei
with the interstellar medium, one can use the different features of
their low--energy spectra 
($ T_{\bar p} \lsim $ 1 GeV, $T_{\bar p}$ being 
the antiproton kinetic--energy). In fact,  
in this energy regime the 
secondary $\bar p$ spectrum
 drops off very markedly because of kinematical reasons, 
while antiprotons due to neutralinos would show a milder fall off.

In Ref. \cite{pierre} the low--energy cosmic--ray $\bar p$ spectrum 
has been calculated, in terms of a 
possible contribution of cosmic--ray antiprotons due
to neutralino annihilation in the galactic halo and of a newly calculated 
flux for secondary antiprotons. The total spectrum has been compared 
with the data collected by the balloon--borne BESS
spectrometer during its flight in 1995 (hereafter
referred to as BESS95 data) \cite{bess}. 
The BESS95 data consist of 43 antiprotons 
grouped in 5 narrow energy windows over the total kinetic--energy range
$0.180 ~ {\rm GeV} \leq  T_{\bar{p}} \leq 1.4 ~{\rm GeV}$. 

In the present paper we evaluate the cosmic--ray antiproton flux as in 
Ref. \cite{pierre}, restricting the supersymmetric configurations to those 
of set $S$ and we 
compare our predictions with the BESS95 data. We refer to \cite{pierre} for all
the details pertaining the evaluation 
of the production of primary and secondary antiprotons 
as well as for the 
properties related to their propagation in the halo
and in the heliosphere. Here we only recall 
the features of the neutralino mass distribution function 
adopted in \cite{pierre} as well as here. This mass distribution function 
is taken spheroidal and parameterized 
as a function $\rho_\chi (r,z)$ of the radial distance $r$
from the galactic center in the galactic plane
and of the vertical distance $z$ from the galactic plane in the form 

\begin{equation}
\rho_\chi (r,z) = \rho_{\chi}\,\, \frac{a^2 + r^2_\odot}{a^2 + r^2 +
z^2/f^2},
\label{eq:mass_DF}
\end{equation}

\noindent
where $a$ is the core radius of the halo,
$r_\odot$ is the distance of the Sun from the
galactic center and $f$ is a parameter which describes the flattening  of the
halo.
Here we take the values: $a=3.5$ kpc, $r_\odot = 8$ kpc. 
The quantity $\rho_{\chi}$ denotes the local value of the neutralino
matter density, which is factorized as 
$\rho_{\chi} = \xi \rho_l$, in terms of the total local dark
matter density $\rho_l$. 
The density $\rho_l$  is calculated by taking into account the
contribution given by the matter density of Eq.(\ref{eq:mass_DF}) to the local
rotational velocity. We use here 
the value  $\rho_l = 0.3$ GeV cm$^{-3}$ in the case of a spherical halo 
($f = 1$). When $f < 1$ (oblate
spheroidal distribution), $\rho_l$ is given by \cite{bt,turner2}
\begin{equation}
\rho_l (f) = \rho_l (f = 1)\, \frac{\sqrt{1-f^2}}{f \,
{\rm Arcsin} \sqrt{1-f^2}}. 
\end{equation}

Our results will be shown for the following values of the local density:
$\rho_l/({\rm GeV cm}^{-3})$ = 0.1 ($f$ = 1), 0.3  ($f$ = 1), 
0.5 ($f$ = 0.50), 0.7 ($f$ = 0.33). For each value of 
$\rho_l$, we will give the top--of--atmosphere (TOA) antiproton fluxes,
as the sum of the secondary flux and of 
the primary flux due to neutralino annihilation for the various 
supersymmetric configurations of set $S$, pertaining 
to that specific value of $\rho_l$.

Fig. 2 displays the scatter plots for the top--of--atmosphere (TOA)  
antiproton fluxes calculated at $T_{\bar p} = 0.24$ GeV, 
 to conform to the energy range
of  the first bin of the BESS95 data (0.175 GeV $\leq T_{\bar p} \leq$ 0.3 GeV).
This figure also shows the band of the experimental results in this bin. 
We find that, while most of the susy configurations of the appropriate 
subset of $S$ stay inside the experimental band for 
$\rho_l$ = 0.1, 0.3 GeV cm$^{-3}$, at higher values of $\rho_l$ a large 
number of configurations provide $\bar p$ fluxes in excess of the 
experimental results. This occurrence is easily understood on the basis 
of the different dependence on $\rho_l$ of the 
direct detection rate and of the $\bar p$ flux, linear in the 
first case and quadratic in the second one. 

To present the  comparison of our theoretical fluxes with the experimental 
ones in a more comprehensive way, we give in Fig. 3 the scatter plots 
of the reduced chi--square
$\chi_{\rm red}^2$ values, obtained for the configurations of 
each of the four subsets of $S$, in a fit to the BESS data over the whole 
energy range (0.175 GeV $\leq T_{\bar p} \leq$ 1.4 GeV). 
In these scatter plots we have included only configurations which are at 
least at the level of the experimental value (within 1--$\sigma$) in 
the first energy bin. 
We remark that, apart from values of $\rho_l$ close to the largest value 
of its physical range, 
a large sample of the susy configurations, relevant for annual modulation 
data in direct search of relic neutralinos, provides a good fit to
the current data on cosmic antiprotons. 

By way of example, we show 
in Fig. 4 a sample of fluxes obtained with a few susy configurations 
of set $S$ which give a good fit to the BESS95 data. 
The parameter $P$, which denotes the various supersymmetric 
configurations, is the gaugino fractional weight 
(for instance, a neutralino which is a pure
gaugino  has $P$ = 1). 

{From} Fig. 3 it also turns out that, especially at large $\rho_l$, many 
configurations of set $S$ have very large values of $\chi_{\rm red}^2$.
Configurations which have $\chi_{\rm red}^2 > 4$ may be considered 
as strongly disfavoured on the basis of the present 
experimental data on cosmic antiprotons ($\chi_{\rm red}^2 = 4$ corresponds
to a 99.9\% C.L. in our case of 5 degrees of freedom).
However, because of the many uncertainties involved in the current 
theoretical evaluations (for instance, effects due to solar modulation 
of the original interstellar fluxes) 
and in the experimental measurements of the antiproton spectrum,
a strict exclusion criterion may be applied only to configurations
which are in large excess of $\chi_{\rm red}^2 = 4$. 
By the same arguments, from our results we derive that  
high values of $\rho_l$, i.e. 
$\rho_l \sim 0.7$ GeV cm$^{-3}$, appear to be disfavoured. 
Hereafter, configurations with $\chi_{\rm red}^2 > 4$ will be denoted by
special symbols.

In Fig. 5 we give $\chi_{\rm red}^2$ versus the neutralino relic abundance. 
On the basis of these results, it turns out that the central values 
of the range for the local density, 0.3 GeV cm$^{-3}$ 
$\lsim \rho_l \lsim 0.5$ GeV cm$^{-3}$, 
is the best suited for neutralinos of cosmological interest. 
The break at $\Omega_\chi h^2 = 0.01$ in the shape of the
scatter plots in this figure is due to the rescaling in
the local density introduced through the parameter $\xi$
for $\Omega_\chi h^2 \leq (\Omega h^2)_{\rm min} \equiv 0.01$.
Because of this, one roughly has: $\Phi_{\bar p} \propto
(\sigma_{\rm ann})^{-1} \propto \Omega_\chi h^2$ for
$\Omega_\chi h^2 \leq (\Omega h^2)_{\rm min}$, and
$\Phi_{\bar p} \propto \sigma_{\rm ann} \propto (\Omega_\chi h^2)^{-1}$ 
otherwise, where $\sigma_{\rm ann}$ is the neutralino--pair annihilation
cross section.

The various results discussed in this section show the 
remarkable property that a number  of the supersymmetric 
configurations singled out by the annual modulation data  may 
indeed produce 
measurable effects in the low--energy part of the $\bar p$ spectrum. 
 Thus the joint use of the 
annual modulation data in direct detection and of the 
measurements of cosmic--ray antiprotons is extremely useful in pinning 
down a number of important properties of relic neutralinos and 
show the 
character of complementarity of these two classes of experimental 
searches for particle dark matter. 
This stresses the great interest for 
the analyses now under way of new antiproton data, those collected 
by a recent balloon flight carried out by the BESS Collaboration 
\cite{orito} and those measured by the AMS experiment \cite{ams} 
during the June 1998 Shuttle flight.

\section{Neutrinos from Earth and Sun}

The signals we are going to discuss now 
consist of
the fluxes of up--going muons in a neutrino telescope, generated by
neutrinos which are produced by pair annihilations
of neutralinos captured and accumulated inside the Earth and the
Sun \cite{flux,mosc,bere2}. The process goes through the following steps: 
a) capture by the celestial body of the relic neutralinos through a slow--down
process due essentially to neutralino elastic scattering off the nuclei of the
macroscopic body,
b)  accumulation of the captured neutralinos in the central part
of the celestial body, c) neutralino--neutralino annihilation with emission
of neutrinos, 
d) propagation of neutrinos and conversion of  their
$\nu_{\mu}$ component into muons in the rock surrounding the detector (or, much
less efficiently, inside the detector), and finally
e) propagation and detection of the ensuing up--going muons in the detector.

The various quantities relevant for the previous
steps are calculated here according to the method described in
Ref. \cite{mosc,bere2}, to which we refer for further details.
Here we only wish to recall that the neutrino fluxes are proportional 
to  the annihilation rate of the neutralinos inside the 
macroscopic body,   ${\Gamma}_A$, which is given by \cite{griest}

\beq
\Gamma_A={C\over 2} {\rm tanh}^2 \left ({t\over \tau_A} \right), 
\label{eq:gamma}
\eeq

\noindent
where $t$ is the age of the macroscopic body ($t \simeq 4.5~{\rm Gyr}$ for Sun
and Earth) and 
$\tau_A = (C C_A)^{-1/2}$;  $C_A$ is the annihilation rate 
proportional to the neutralino--neutralino annihilation cross--section and 
$C$ is the capture rate, proportional to the neutralino--nucleus 
cross--section and to the neutralino local density. 
{From} Eq.(\ref{eq:gamma}) it follows that in a given macroscopic body the
equilibrium between capture and annihilation ({\it i.e.}
$\Gamma_A \sim C/2$ ) is established
only when $t \gsim \tau_A$.

{From} the evaluation of the 
annihilation rate for neutralinos inside the Earth
and the Sun it turns out that,
for the Earth, the equilibrium condition depends critically on the
values of the model parameters, and is not realized in wide regions of
the parameter space. Consequently, for these regions the signal due to
neutralino annihilation may be significantly attenuated. 
Capture by the Earth is very efficient for $m_\chi \lsim 70$ GeV because of 
mass--matching condition between the neutralino mass and the nuclear mass of 
some important chemical constituents of the Earth (O, Si, Mg, Fe). 
On the contrary,
in the case of the Sun, the capture--annihilation equilibrium is
reached for the whole range of
$m_\chi$, due to the much more efficient capture rate due to the
stronger gravitational field \cite{gould}.

Let us now report our results. In Fig. 6 we display the 
scatter plots for the flux of the up--going muons from the center of the 
Earth, for various values of the  local dark matter density 
$\rho_l$. In this figure is also reported the current $90\%$ C.L. 
experimental upper 
bound on $\Phi_{\mu}^{\rm Earth}$, obtained by MACRO \cite{macro} (a similar
experimental upper bound is given in Ref. \cite{baksan}). 
We note the particular enhancement at 
$m_{\chi} \sim (50 - 60)$ GeV, due to the mass--matching 
effect. It is also noticeable an effect of suppression and spreading 
of the fluxes for $m_{\chi} \gsim 70$ GeV at  
$\rho_l \gsim 0.3$ GeV cm$^{-3}$. 
This is due to the fact that the configurations with these 
large values of $\rho_l$ may imply, because of the annual modulation data,
a neutralino--nucleus cross--section 
which is too small to establish an efficient capture rate, necessary for 
the capture-annihilation 
equilibrium in Earth. This is shown in Fig. 7, whose 
scatter plots show that indeed in the case of large values of 
$\rho_l$ many configurations induce sizeable deviations from the 
value $\tanh^2 (t/t_A) = 1$, which guarantees equilibrium in the 
macroscopic body. 

In Fig. 6  the configurations which would be 
excluded on the basis of the antiproton data, as previously discussed in 
Sect. II, are denoted differently from those which would survive this 
criterion. 

By comparing our scatter plots with the experimental MACRO upper limit, 
one notices that a number of supersymmetric configurations provide 
a flux in excess of this experimental bound and might then be considered
as excluded. However, 
it has to be recalled that a possible neutrino oscillation effect 
\cite{superkamioka} may 
be operative here and  affect   the indirect neutralino signal as 
well as the background consisting in atmospheric neutrinos.  
 The influence of a neutrino oscillation on the problem under 
discussion would depend very critically on the nature of the 
oscillation itself ($\nu_{\mu}$ converting either into a $\nu_{\tau}$ or 
into a sterile neutrino). Therefore a strict enforcement of the
current upper bound on $\Phi_\mu^{\rm Earth}$ should be applied with caution
as long as the neutrino oscillation properties are not fully understood.  
However, it is rewarding that the set $S$ of 
supersymmetric configurations is quite accessible to relic neutralino 
indirect search by measurements of up--going fluxes, as displayed in our 
results of Fig. 6. 

Fig. 8 displays the up--going muon fluxes from the Sun. We notice that 
a number of configurations of higher masses are easily accessible to the
measurements of these fluxes.

\section{Conclusions}

We have proved that a sizeable fraction of the 
supersymmetric neutralino configurations singled out by the 
DAMA/NaI data on a possible  annual modulation effect in WIMP direct search,  
may provide signals detectable by two of the most 
interesting methods  of indirect search for relic particles: 
measurement of cosmic--ray antiprotons and 
measurement of neutrino fluxes from  Earth and  Sun. 

For the case of cosmic antiprotons, it has been shown that 
present data are well fitted by total spectra which include a $\bar p$
contribution from neutralino--pair annihilation, with neutralino 
configurations which are relevant for annual modulation in direct detection. 
These data can also be used to reduce the total sample of the
supersymmetric configurations under study, and to narrow the range 
of the local density, by disfavouring its largest values.
Investigation by measurements of cosmic $\bar p$'s looks very promising 
in view of the collections and analyses of more statistically 
significant sets of data in the 
low--energy regime which are currently under way \cite{orito,ams} and
which may soon provide further relevant information.

Measurements of neutrino fluxes from Earth and Sun, due to capture and annihilation
of neutralinos inside these celestial bodies, have been proved to be sensitive
to neutralino configurations singled out by the annual modulation data. An
appropriate interpretation
of these measurements preliminarily requires some clarification of the 
oscillation neutrino properties. 

In conclusion, we have shown how  
the measurements of cosmic--ray antiprotons and of neutrino fluxes from Earth
and Sun are complementary to those of direct detection in investigating
astrophysical and cosmological properties of the relic neutralinos which 
may be at the origin of the possible modulation effect in direct search.

\vfill
\eject

\begin{center}
{\large FIGURE CAPTIONS}
\end{center}

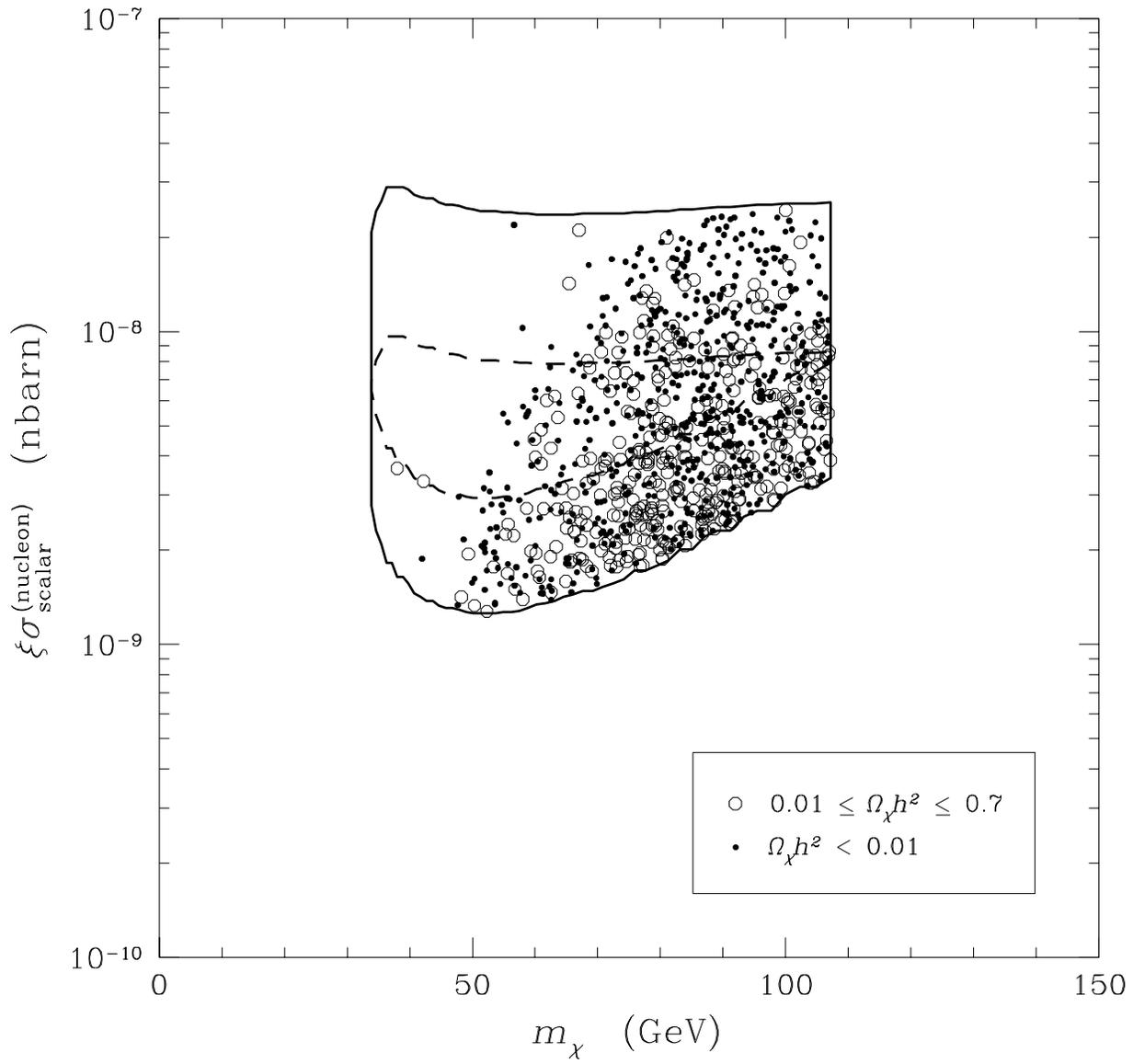
{\bf Figure 1} -- Scatter plot of set $S$ in the plane 
$m_{\chi}$--$\xi \sigma_{\rm scalar}^{(\rm nucleon)}$. 
The dashed contour line
delimits the 2--$\sigma$ C.L. region, obtained by the DAMA/NaI Collaboration, 
 by combining together the data of
the two running periods of the annual modulation experiment \cite{dama2}.  
The solid contour  line is obtained from the dashed line, which refers to the
value $\rho_l = 0.3 ~$ GeV cm$^{-3}$, by accounting for 
the uncertainty range of $\rho_l$, as explained in Sect. I (the region 
delimited by the solid line is denoted as region $R$ in the text). 
Displayed in this figure are only the representative points of the susy 
parameter space, which fall inside the region $R$ (this set of configurations
is defined as set $S$).
Dots and circles denote neutralino configurations depending on their values 
for the neutralino relic abundance.

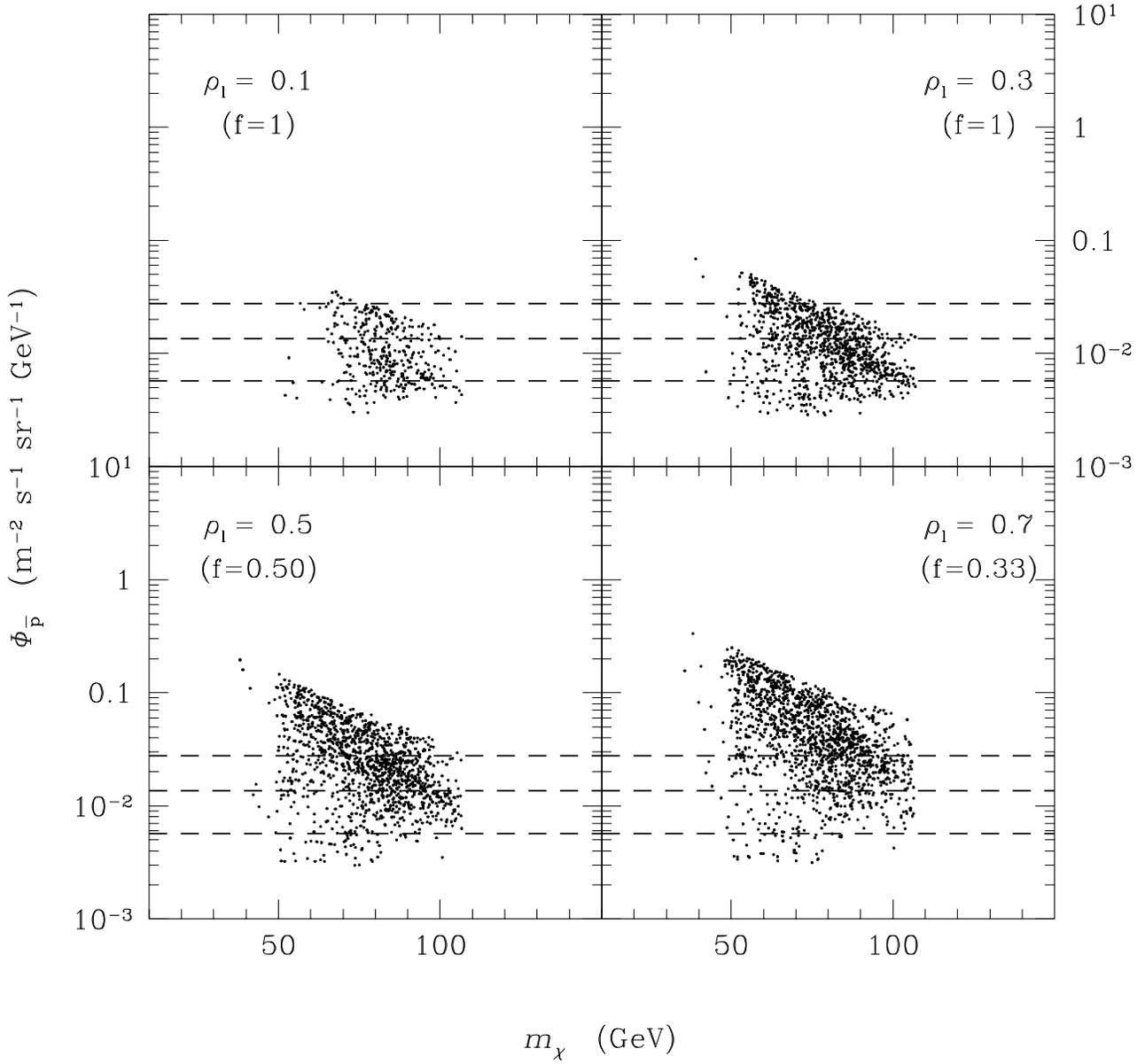
{\bf Figure 2} -- 
 Scatter plots for the TOA antiproton fluxes calculated at 
$T_{\bar p} = 0.24$ GeV versus the neutralino mass. 
The configurations of set $S$ are 
subdivided into the 4 panels, depending on the corresponding value of 
the local density:
$\rho_l/({\rm GeV cm}^{-3})$ = 0.1 ($f$ = 1), 0.3  ($f$ = 1), 
0.5 ($f$ = 0.50), 0.7 ($f$ = 0.33). The 
 dashed lines denote the central value and the 1--$\sigma$ band of the 
BESS95 data in the first energy bin: 
0.175 GeV $\leq T_{\bar p} \leq$ 0.3 GeV. 

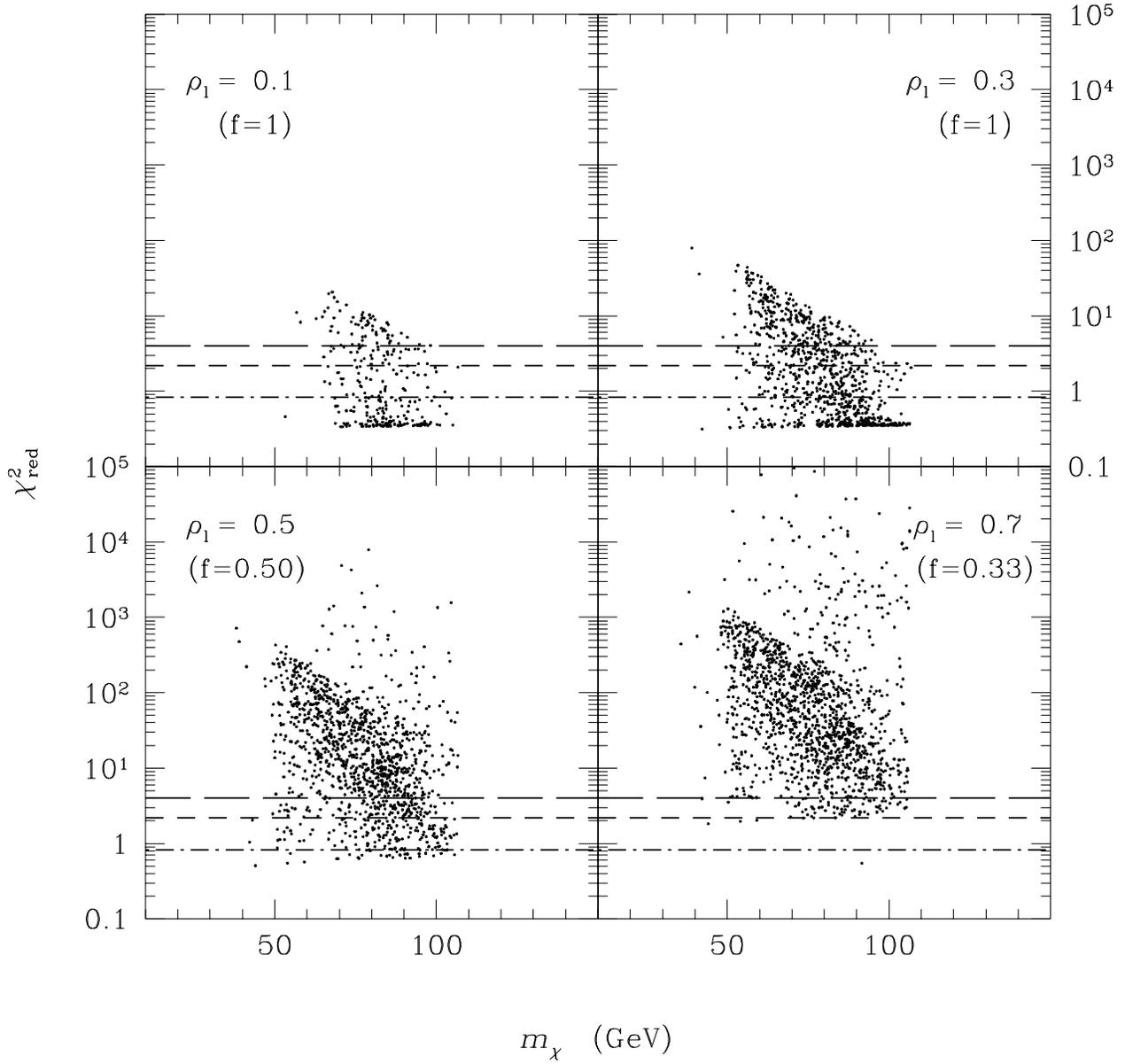
{\bf Figure 3} -- 
Scatter plots 
of the  $\chi_{\rm red}^2$ values, obtained for the configurations of 
each of the four subsets of $S$, in a fit to the BESS95 data over the whole 
energy range (0.175 GeV $\leq T_{\bar p} \leq$ 1.4 GeV) versus the neutralino 
mass. 
Only configurations which are at 
least at the level of the experimental value (within 1--$\sigma$) in 
the first energy bin are included. 
The three 
horizontal lines represent: the $\chi_{\rm red}^2$ 
in a fit of the data using the secondary flux only (dot--dashed line), 
and the $\chi_{\rm red}^2$ in a  fit using the total 
(primary plus secondary) flux at $95\%$ C.L. (short--dashed line) and 
at $99.9\%$ C.L. (long--dashed line). The $\rho_l$ values are in units
of GeV cm$^{-3}$.

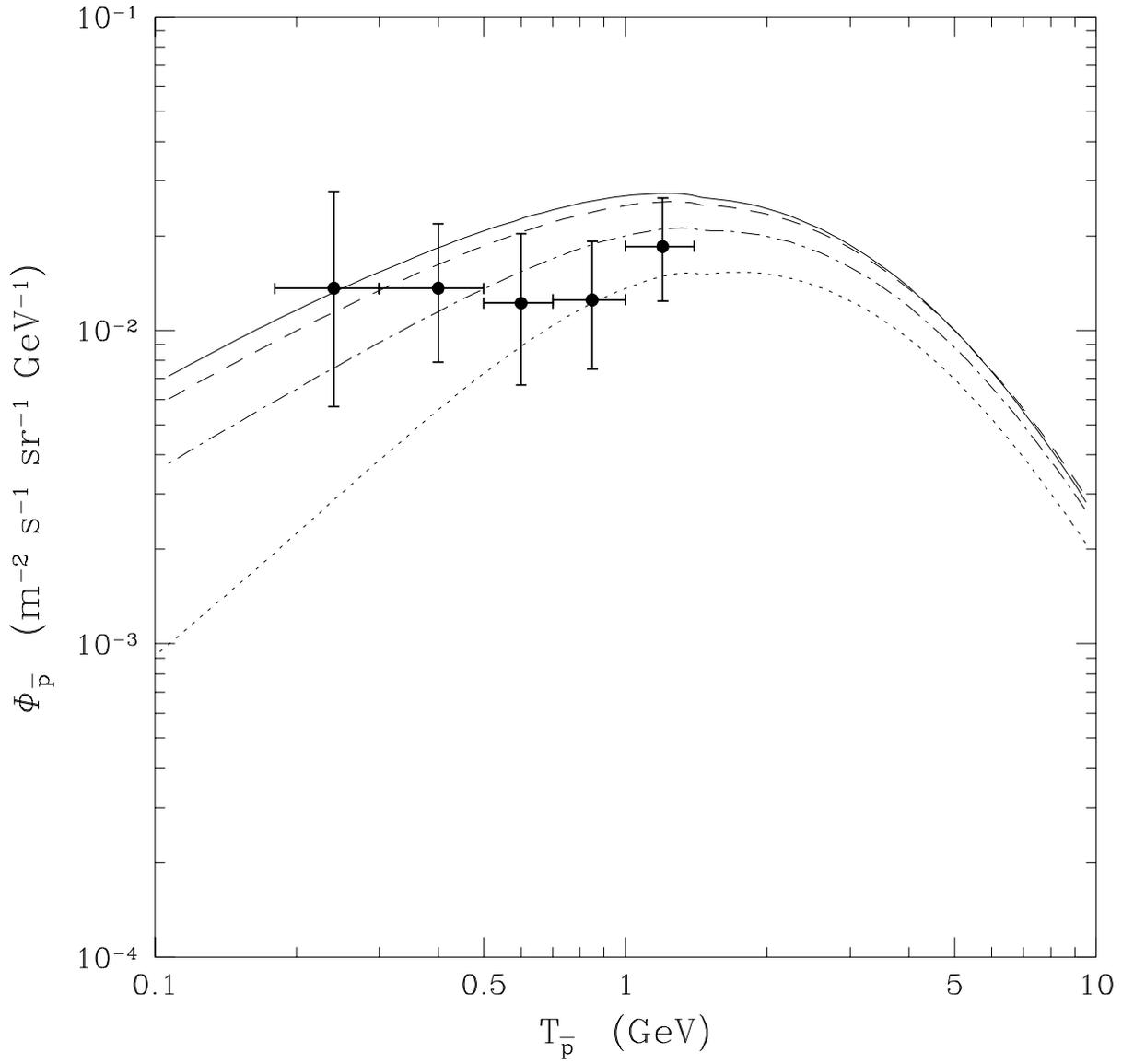
{\bf Figure 4} -- 
Some examples of TOA fluxes which provide a good fit to the BESS95 data,
for $\rho_l = 0.3$ GeV cm$^{-3}$.
The dotted line denotes the secondary antiproton flux only.
The solid, dashed and dot--dashed lines denote the total
antiproton fluxes, obtained as a sum of the secondary spectrum and
of the primary spectra due, respectively, to the following neutralino
configurations: 
$m_\chi = 62$ GeV, $\Omega_\chi h^2 = 0.03$ and $P=0.98$ (solid line);
$m_\chi = 81$ GeV, $\Omega_\chi h^2 = 0.02$ and $P=0.99$ (dashed line);
$m_\chi = 95$ GeV, $\Omega_\chi h^2 = 0.03$ and $P=0.99$ (dot--dashed line).

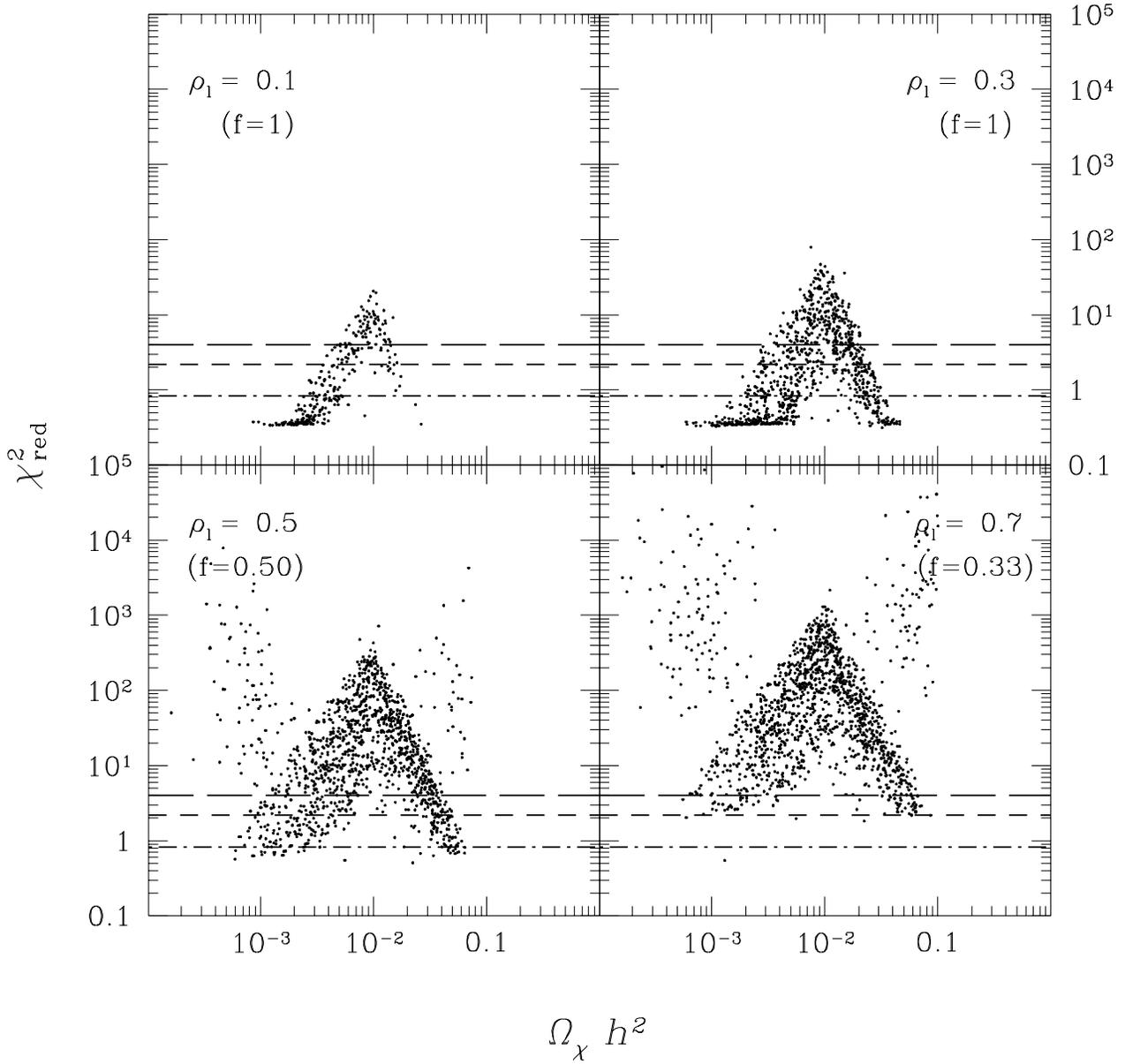
{\bf Figure 5} --
Scatter plots 
of the  $\chi_{\rm red}^2$ values, obtained for the configurations of 
each of the four subsets of $S$, in a fit to the BESS data over the whole 
energy range (0.175 GeV $\leq T_{\bar p} \leq$ 1.4 GeV) versus the neutralino 
relic abundance. Notations as in Fig. 3. 

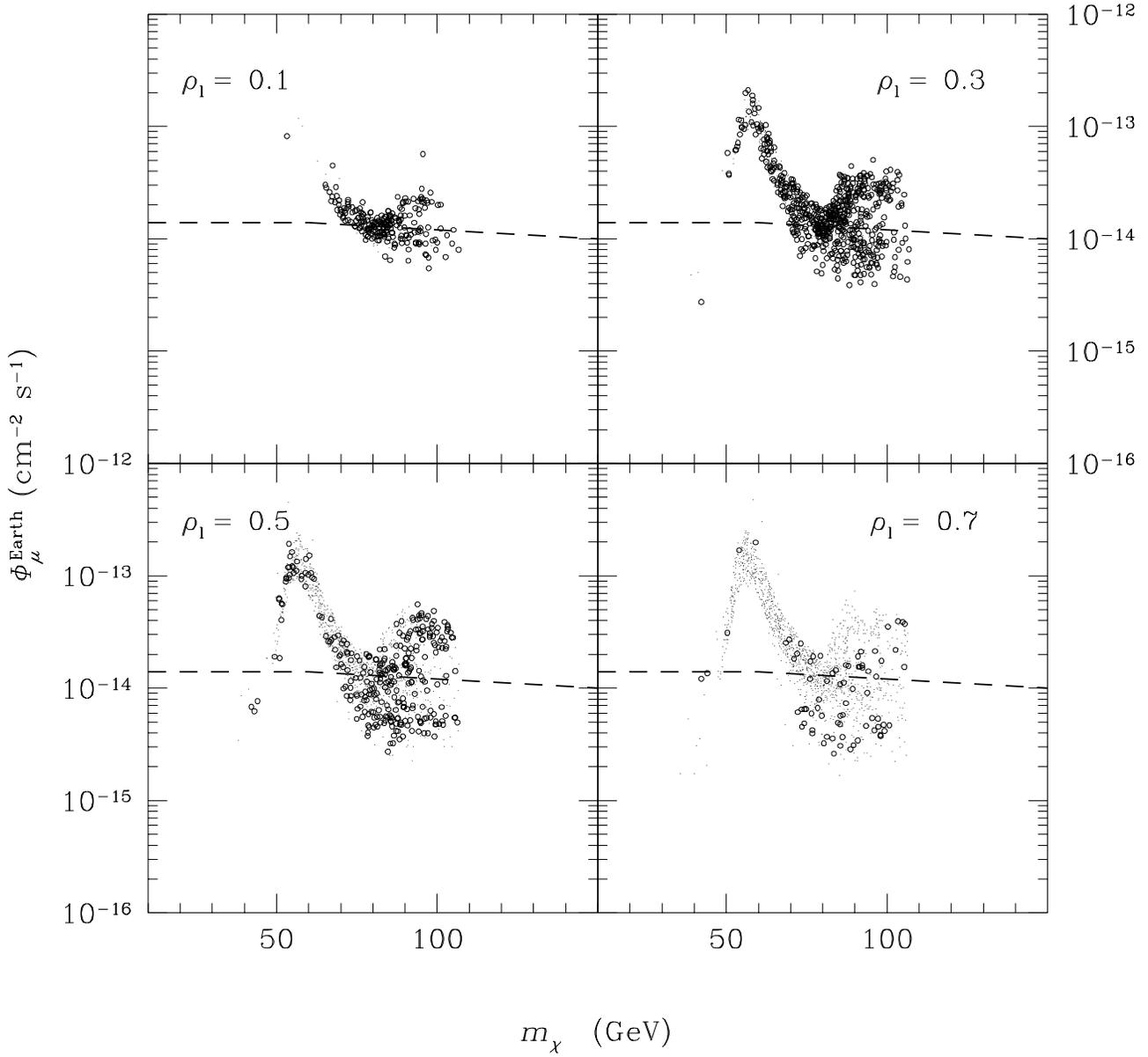
{\bf Figure 6}
Scatter plots for the up--going muon fluxes from the center of the Earth versus
the neutralino mass.
The configurations of set $S$ are 
subdivided into the 4 panels, depending on the corresponding value 
the local density:\\
$\rho_l/({\rm GeV cm}^{-3})$ = 0.1 ($f$ = 1), 0.3  ($f$ = 1), 
0.5 ($f$ = 0.50), 0.7 ($f$ = 0.33). 
Dots denote configurations which could be excluded on the basis of the BESS95 
antiproton data, circles denote configurations which survive this exclusion
criterion. The dashed line denotes the MACRO upper bound. 

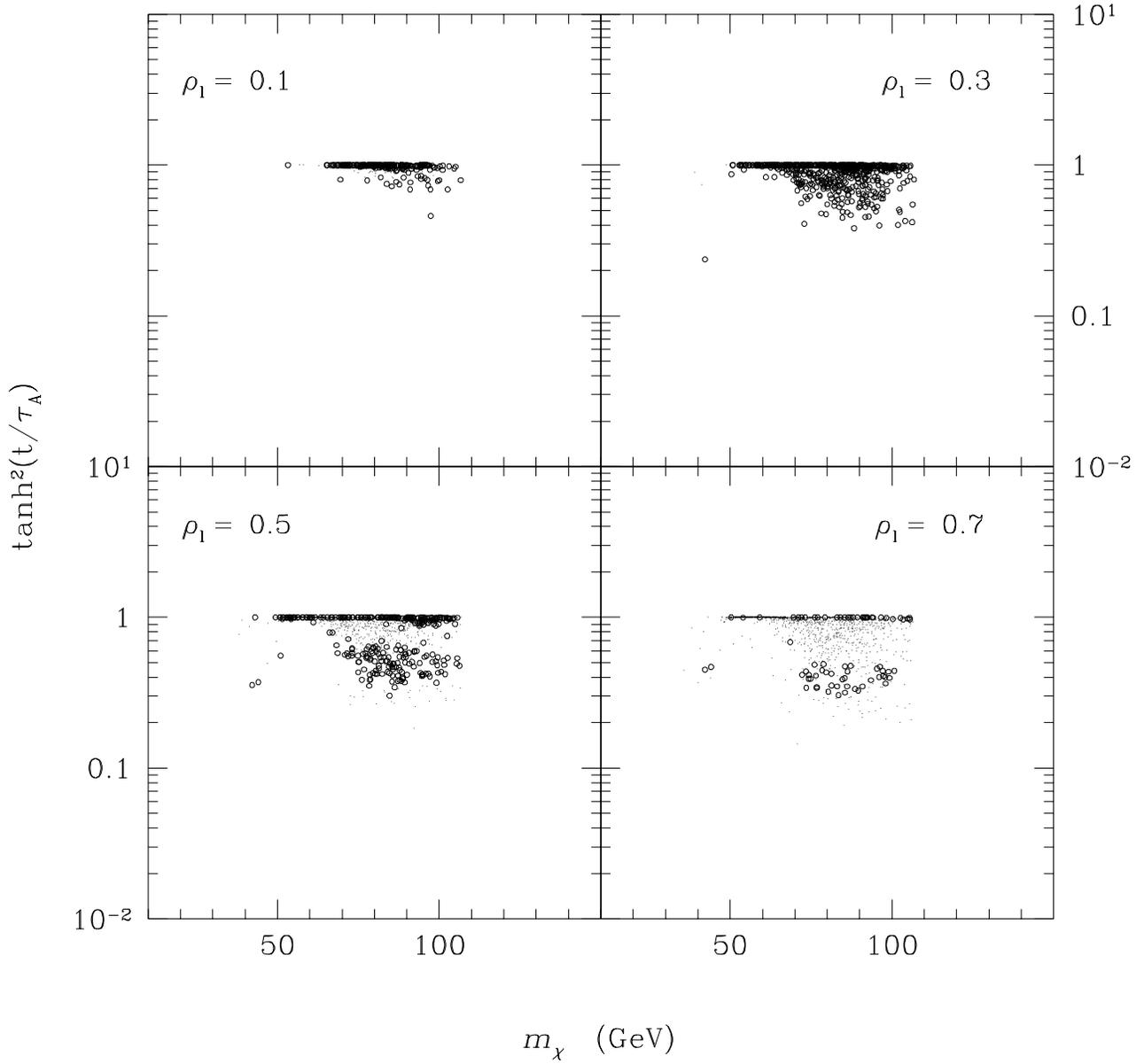
{\bf Figure 7} -- 
Scatter plots of the quantity ${\rm tanh}^2 (t/ \tau_A)$ 
versus the neutralino mass. Other specifications are as in Fig. 6. 

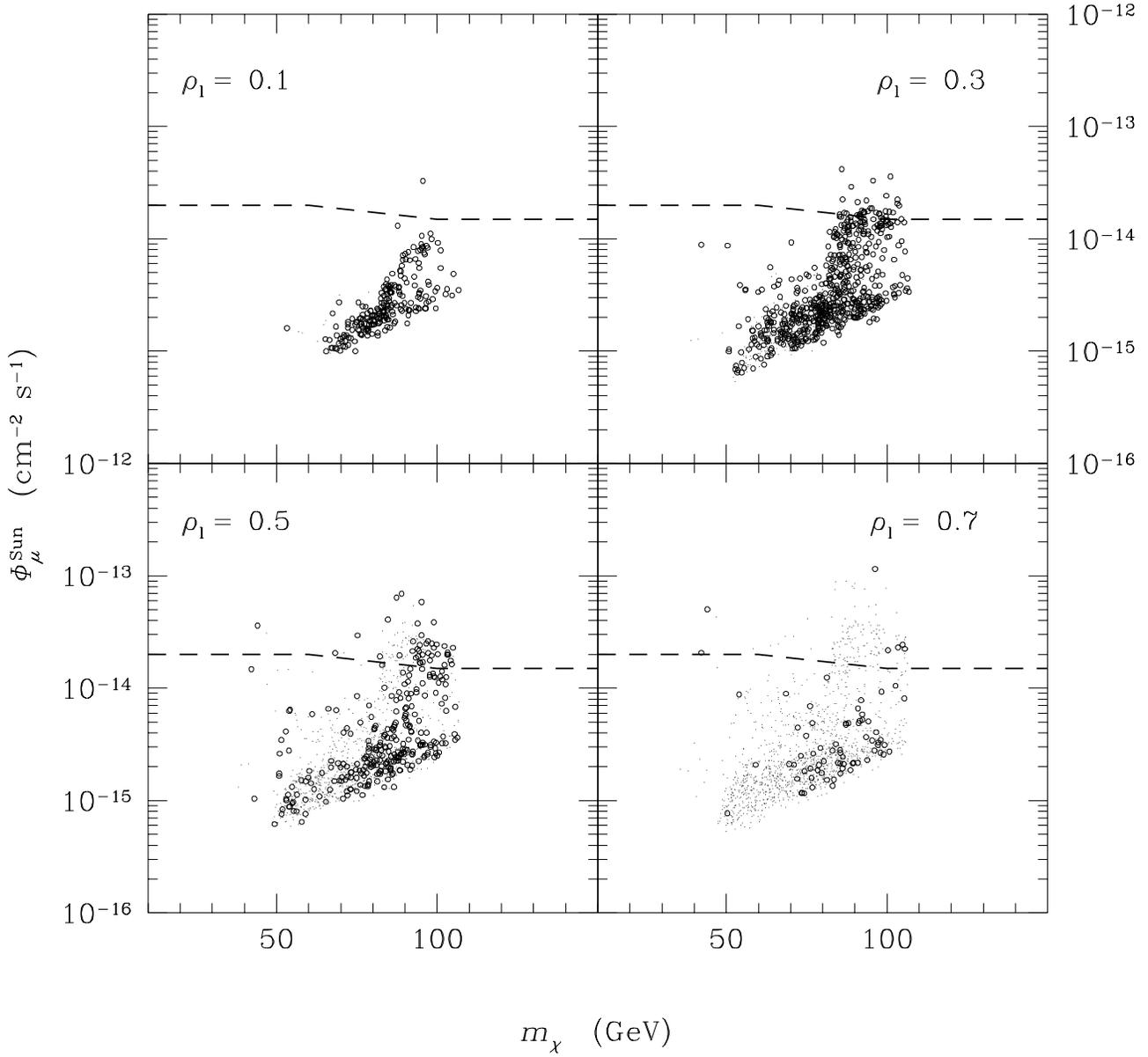
{\bf Figure 8} -- 
Scatter plots for the up--going muon fluxes from the  Sun versus
the neutralino mass. Other specifications are as in Fig. 6.

\vfill\eject


\begin{thebibliography}{99}


\bibitem{noi0} A. Bottino, F. Donato, N. Fornengo and S. Scopel, 
Phys. Lett. B423 (1998) 109.

 
\bibitem{noi1} A. Bottino, F. Donato, N. Fornengo and S. Scopel, 
            Torino University preprint DFTT 41/98, hep--ph/9808456.

\bibitem{noi2} A. Bottino, F. Donato, N. Fornengo and S. Scopel, 
            Torino University preprint DFTT 48/98, hep--ph/9808459.


\bibitem{dama1} R. Bernabei et al., Phys. Lett. B424 (1998) 195.


\bibitem{dama2}      R. Bernabei et al., Roma II University preprints: 
                     ROMA2F/98/27 and ROMA2F/98/34, August 1998.


\bibitem{turner1}   E. Gates, G. Gyuk and M.S. Turner,
                     Phys. Rev. Lett. {\bf 74}, 3724 (1995);
                     Phys. Rev. {\bf D53}, 4138 (1996).

\bibitem{turner2}
E. Gates, G. Gyuk and M.S. Turner,
  Astrophys. J. Lett. {\bf 449}, L123 (1995). 


\bibitem{turner}    E. Gates, G. Gyuk and M.S. Turner,
                     talk presented at {\it 18th Texas Symposium on 
                     Relativistic Astrophysics},
                     Chicago, December 1996, astro--ph/9704253.





\bibitem{pbar} J. Silk and M. Srednicki, Phys. Rev. Lett. {\bf 53}, 624 (1984);
J. Ellis, R.A. Flores, K. Freese, S. Ritz, D. Seckel and J. Silk, Phys. Lett.
{\bf B214}, 403 (1988); F. Stecker, S. Rudaz and T. Walsch, Phys.
Rev. Lett. {\bf 55}, 2622 (1985);
J.S. Hagelin and G.L. Kane, Nucl. Phys. {\bf B263}, 399 (1986);
S. Rudaz and F.W. Stecker, Astrophys. J. {\bf 325}, 16 (1988);
F. Stecker and A. Tylka, Astrophys. J. {\bf 336}, L51 (1989);
G. Jungman and M. Kamionkowski, Phys. Rev. {\bf D49}, 2316 (1994);
T. Mitsui, K. Maki and S. Orito, Phys. Lett. {\bf B389}, 169
(1996).


\bibitem{pbarnostro} A. Bottino, C. Favero, N. Fornengo and G. Mignola, Astropart.
Phys. {\bf 3}, 77 (1995).

\bibitem{pierre} A. Bottino, F. Donato, N. Fornengo and P. Salati, 
astro--ph/9804137 (to appear in Phys. Rev. D)

\bibitem{flux} 
J. Silk, K. Olive and M. Srednicki, Phys. Rev. Lett. 55 (1985) 
257; T. Gaisser, G. Steigman and S. Tilav, Phys. Rev. D34 (1986) 2206;  K.
Freese, Phys. Lett. B167 (1986) 295; K. Griest and S. Seckel, Nucl. Phys. B279 
(1987) 804; 
G.F. Giudice and E. Roulet, {\it Nucl. Phys.} {\bf B316}
(1989) 429; 
G.B. Gelmini, P. Gondolo and E. Roulet, {\it Nucl. Phys.}
{\bf B351} (1991) 623; 
M. Kamionkowski,  {\it Phys. Rev.} {\bf D44} (1991) 3021; 
A. Bottino, V. de Alfaro, N. Fornengo, G. Mignola and
M. Pignone, {\it Phys. Lett.} {\bf B265} (1991) 57; 
F. Halzen, M. Kamionkowski and T. Steltzer, {\it Phys. Rev.} {\bf D45}
(1992) 4439; 
V.S. Berezinsky, {\it Nucl. Phys.} (Proc. Suppl.) {\bf B31} (1993) 413
(Proc. Neutrino 92, Ed. A. Morales) 
M. Mori et al., {\it Phys. Rev.} {\bf D48} (1993) 5505; 
M. Drees, G. Jungman, M. Kamionkowski and M.M. Nojiri,
{\it Phys. Rev.} {\bf D49}
(1994) 636; 
R. Gandhi, J.L. Lopez, D.V. Nanopoulos, K. Yuan and A. Zichichi,
{\it Phys. Rev.} {\bf D49} (1994) 3691; 
L. Bergstr\"om, J. Edsj\"o and P. Gondolo, Phys. Rev. D55 (1997) 1765. 
L. Bergstr\"om, J. Edsj\"o and M. Kamionkowski, Astrop. Phys. 7 (1997) 147.

\bibitem{mosc} A. Bottino, N. Fornengo, G. Mignola and L. Moscoso,
{\em Astroparticle Physics} {\bf 3} (1995) 65.


\bibitem{bere2} V. Berezinsky, A. Bottino, J. Ellis, N. Fornengo,
G. Mignola and S. Scopel, Astrop. Phys. 5 (1966) 333.

\bibitem{ouromega}   A. Bottino, V. de Alfaro, N. Fornengo, G. Mignola 
                     and M. Pignone,
                     Astropart. Phys. {\bf 2}, 67 (1994).



\bibitem{omegamatter} W.L. Freedman, R. Kirshner and C. Lineweaver, 
                    talks given at {\it International Conference of 
                    Cosmology and Particle Physics}, Geneva, June 1998,
                     http://wwwth.cern.ch/capp98/programme.html; 
                    M. White, astro--ph/9802295; 
                    N.A. Bahcall and X. Fan, astro--ph/9804082;
                    C. Lineweaver, astro--ph/9805326.


\bibitem{bess} H. Matsunaga {\it et al.} (BESS Collaboration), in
            {\it Proceedings of the 25th International Conference of Cosmic
                Rays, Durban, 1997} and A. Moiseev, private communication. 


\bibitem{bt} J. Binney and S. Tremaine, {\it Galactic Dynamics}
            (Princeton University Press, 1987).

\bibitem{griest}
K. Griest and D. Seckel,  {\it Nucl. Phys.} {\bf B283} (1987) 681.


\bibitem{gould}
A. Gould, {\it Ap. J.} {\bf 321} (1987) 571;
{\it Ap. J.} {\bf 328} (1988) 919;
{\it Ap. J.} {\bf 368} (1991) 610.


\bibitem{macro} M. Ambrosio et al. (MACRO Collaboration), MACRO/PUB1/98; 
                T.  Montaruli et al., {\it Proceedings of TAUP97}, Nucl. Phys. 
                {\bf B70} (Proc. Suppl.), 367 (1999).


\bibitem{baksan} M.M. Boliev et al., {\it Proceedings of TAUP97}, Nucl. Phys.
                {\bf B70} (Proc. Suppl.), 371 (1999).

\bibitem{superkamioka} Y. Fukuda et al. (Super--Kamiokande Collaboration), 
hep--ex/9807003. 

\bibitem{orito} S. Orito, talk at International Conference on High 
Energy Physics, Vancouver, July 1998.

\bibitem{ams} S. Ahlen at al., Nucl. Instrum. Methods A350 (1994) 351. 


\bibitem{damapsa}    R. Bernabei et al., Phys. Lett. {\bf B389}, 757 (1996).

\end{thebibliography}
\end{document}